\documentclass[prd,showpacs,preprintnumbers,twocolumn,amsmath,nofootinbib,amssymb]{revtex4}
\usepackage{graphicx,color,dcolumn,booktabs,bm}
\usepackage{longtable,lscape}
\usepackage{txfonts}
\usepackage{overpic}
\usepackage{float}
\usepackage{amssymb}
\usepackage{epstopdf}
\usepackage{indentfirst}
\usepackage{feynmf}   
\usepackage{slashed}  
\usepackage{cases}
\usepackage{ulem}
\usepackage{color}
\usepackage{float}
\usepackage{multirow}
\usepackage{tikz}
\usepackage{epstopdf}
\usepackage{epsfig,dsfont,amssymb,amsmath,amsfonts,amsbsy,mathrsfs}

\usepackage[colorlinks, citecolor=blue,anchorcolor=red,menucolor=red, linkcolor=red,filecolor=red,runcolor=red,urlcolor=blue,frenchlinks=red]{hyperref}
\makeatletter
\@addtoreset{equation}{section}
\makeatother

\newcommand{\nc}{\newcommand}
\nc{\tj}[1]{\textcolor{red}{Tianjie: #1}}
\allowdisplaybreaks
\begin{document}
\title{Study of the $\omega$ meson family and newly observed $\omega$-like state $X(2240)$}
\author{Cheng-Qun Pang$^{1,2}$\footnote{Corresponding author}}\email{pcq@qhnu.edu.cn}
\author{Ya-Rong Wang$^{1,2}$}
\author{Jing-Fu Hu$^{1,2}$}
\author{Tian-Jie Zhang$^{3}$}
\author{Xiang Liu$^{2,4,5}$\footnote{Corresponding author}}\email{xiangliu@lzu.edu.cn}
\affiliation{
$^1$ College of Physics and Electronic Information, Qinghai Normal University, Xining 810000, China\\
$^2$ Joint Research Center for Physics, Lanzhou University and Qinghai Normal University, Xining 810000, China\\
$^3$ School of Mathematics and Statistics, Ningxia University, Yinchuan 750000, China\\
$^4$School of Physical Science and Technology, Lanzhou University, Lanzhou 730000, China\\
$^5$Research Center for Hadron and CSR Physics, Lanzhou University
and Institute of Modern Physics of CAS, Lanzhou 730000, China}

\begin{abstract}
Since the present $\omega$ meson family has not been established, in this work, we carry out an investigation of {the} mass spectrum and  {Okubo-Zweig-Iizuka a} allowed two-body strong decay of {the} $S$-wave and {the} $D$-wave $\omega$ mesons, and make the comparison with the experimental data of these reported $\omega$ states and {the} $\omega$-like $X(2240)$ state observed by BESIII. By this study, we not only suggest the possible {assignments} to these observed $\omega$ states under the framework of the $\omega$ meson family, but also predict three $\omega$ mesons {($\omega(5S)$, $\omega(2D)$ , and $\omega(4D)$)} which are still missing in experiment. The present study may provide valuable information to further construct the $\omega$ meson family. Considering the present running status of BESIII, we also suggest that BESIII should pay more {attention} to the issue of $\omega$ meson with accumulating more data.
\end{abstract}
\maketitle

\section{introduction}\label{s1}

\begin{figure*}[htbp]
\begin{overpic}[scale=0.36,tics=20]{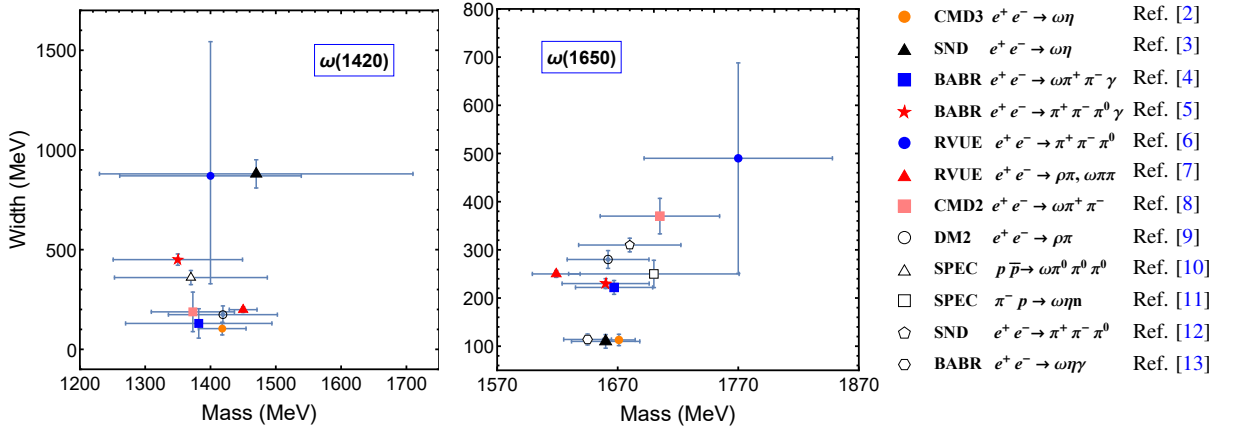}%
\put(100.8,36.4){\footnotesize Ref. \cite{CMD-3:2017tgb}}
\put(100.8,33.572){\footnotesize Ref. \cite{Achasov:2016qvd}}
\put(100.8,30.642){\footnotesize Ref. \cite{Aubert:2007ef}}
\put(100.8,27.8161){\footnotesize Ref. \cite{Aubert:2004kj}}
\put(100.8,24.9904){\footnotesize Ref. \cite{Achasov:2003ir}}
\put(100.8,22.2696){\footnotesize Ref. \cite{Henner:2002iv}}
\put(100.8,19.3388){\footnotesize Ref. \cite{Akhmetshin:2000wv}}
\put(100.8,16.4081){\footnotesize Ref. \cite{Antonelli:1992jx}}
\put(100.8,13.5825){\footnotesize Ref. \cite{Anisovich:2000jx}}
\put(100.8,10.7525){\footnotesize Ref. \cite{Eugenio:2000rf}}
\put(100.8,7.8258){\footnotesize Ref. \cite{Aulchenko:2015mwt}}
\put(100.8,5){\footnotesize Ref. \cite{Aubert:2006jq}}
\end{overpic}
\caption{{\color{black} The} comparisons of the resonance parameters of $\omega(1420)$ and  {$\omega(1650)$}  measured by different experiments. 
 }
 \label{omegapar}
\end{figure*}

As a {well-known} established meson below 1 GeV, $\omega(782)$ is the ground state of the $\omega$ meson family. If you do not carefully read Particle Data Book from Particle Data Group (PDG) \cite{Tanabashi:2018oca}, you must intuitively think that {the} higher states of  the  $\omega$ family {have} be established in experiment at least for these lower  excitations. In fact, there exists a {\color{black}messy} situation for the $\omega$ family as shown in Fig. \ref{omegapar}. {Checking} the PDG data, we find that the {\color{black}inconsistency of resonance parameters for} the most of the  $\omega$ mesons popularly exists in these reported experiment data from different groups, especially for {the} width measurement. This problem did not {\color{black}attract  attention} from the community. In Fig. \ref{omegapar}, we illustrate this {\color{black}messy} situation by listing experimental data, which inspires our interest in further studying this research topic.

We also notice a recent result reported by the BESIII Collaboration. By studying the cross section of the $e^+e^-\to K^+K^-$ process, a resonance structure exists in the $K\bar{K}$ invariant mass spectrum which has {the mass of $M=2239.2\pm7.1\pm11.3$ MeV and the width of} $\Gamma=139.8\pm12.3\pm20.6$ MeV \cite{Ablikim:2018iyx}.  In this work, we tentatively name this structure as $X(2240)$. Although it is treated as the  {\color{black}candidate of  tetraquark} states in Refs. \cite{Lu:2019ira,Azizi:2019ecm}, the properties of $X(2240)$ still remain unclear. 
 Since $X(2240)$ is from the $e^+e^-$ annihilation, its $J^{P}$ must be $1^-$ while its isospin can be either 0 or 1. This experimental phenomenon also indicates that BESIII will be a good platform to study {the} higher states of {the} $\omega$ meson family.
In this work, we will put $X(2240)$ and the other $\omega$ states together when discussing how to construct the $\omega$ meson family.


Among these light mesons, these isovector vector light mesons form the $\rho$ meson family while these isoscalar vector {\color{black}light-mesons} construct the $\omega$ meson family. In fact, the $\omega$ meson family is similar to the $\rho$ meson family which is directly reflected by the fact that $\omega(782)$ and $\rho(770)$ is approximately degenerate in mass. Thus, {the} $\rho$ meson spectrum can be {\color{black}a} scaling point when theoretically constructing the $\omega$ meson spectrum. It is necessary to specify that we assume $\omega$ mesons do not have the $s \bar{s}$ component in this work.

In 2013, {\color{black}the} Lanzhou group systematically studied the mass spectrum and decay behavior of these higher excited $\rho$ mesons \cite{He:2013ttg}, and found two Regge trajectories: [$\rho(770)$, $\rho(1450)$, $\rho(1900)$, $\rho(2150)$] and [$\rho(1700)$, $\rho(2000)$, $\rho(2270)$],  {in which}  the former one and the later one correspond to {the} $S$-wave and $D$-wave states in the $\rho$ meson family, respectively. In the Regge trajectory, the radial quantum number is increasing with {\color{black}the} mass change of these states. We need to explain that these $\rho$ states are from PDG data \cite{Tanabashi:2018oca}.

{ {\color{black}}Because of} the similarity between the $\omega$ meson family and $\rho$ meson family, we can give the Regge trajectories for the discussed $\omega$ meson family {by  the experience of studying the} $\rho$ meson family \cite{He:2013ttg}. In Sec. \ref{rta}, we present the  corresponding analysis of {\color{black}the} Regge trajectory. According to our investigation, we may find that these reported $\omega$-like states  can be categorized into {the} $\omega$ meson family when performing
a simple mass spectrum analysis. Obviously, {the} mass spectrum information is limited to {\color{black}identifying} these possible assignments. Thus, we need to further carry out the calculation of two-body {\color{black}Okubo-Zweig-Lizukan (OZI)} allowed strong decays of them, by which we can distinguish dominant, main and subordinate decay modes of these discussed $\omega$ states. Getting their OZI allowed two-body decay behavior is a key step  {to further clarify} the {\color{black}messy} situation of these $\omega$ states, which will be  {  the} main task of this work.

This paper is organized as follows. After { the {\color{black}Introduction}}, we provide a Regge trajectory analysis for the discussed $\omega$ meson family in Sec. \ref{rta}. And then, we further calculate the partial decay widths of these OZI allowed two-body  decays, where the quark pair creation (QPC) model is adopted. By these obtained results, we further select their dominant, main and subordinate decay modes, and some typical ratios of decays (see Sec. \ref{s2}). The paper ends with a short summary in Sec. \ref{s3}.

\section{Regge trajectory analysis}\label{rta}
{{Finding the Regge trajectory} is an effective approach to study a light-meson spectrum \cite{Chew:1962eu,Anisovich:2000kxa}}.
 {The masses and radial quantum numbers of the light mesons in the same meson family satisfy the following relation:}
\begin{eqnarray}
M^2=M_0^2+(n-1)\mu^2,\label{rt}
\end{eqnarray}
where {$M_0$} represents the mass of {the }ground state. In addition, $\mu^2$ is the
trajectory slope and $n$ is the radial quantum number of the corresponding meson with {the} mass $M$.

\begin{figure*}[htbp]
\centering%
\scalebox{0.70}{\includegraphics{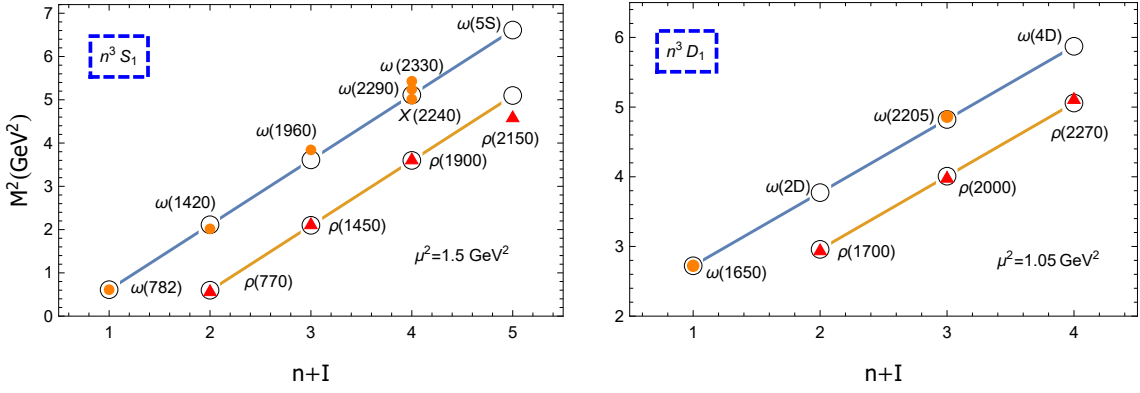}}\\
\caption{{\color{black} The}  Regge trajectories of $\omega$ and $\rho$
states. {{Here I is the isospin of mesons. {{The }}open circle and the filled geometry (circle and triangle) are the theoretical and experimental values, respectively}}.
\label{regge} }
\end{figure*}

Adopting Regge trajectory analysis and combining with the $\rho$ meson spectrum, we may construct two Regge trajectories of the $\omega$ meson family.
In Fig. \ref{regge}, we plot the Regge trajectories of the $\rho$ meson according to the information in Ref. \cite{He:2013ttg}, where $\mu^2=1.5$ GeV$^2$ for the $S$-wave states and $\mu^2=1.05$ GeV$^2$ for the $D$-wave states. Then, by taking ground states $\omega(782)$ and $\omega(1650)$ as input and adopting almost the same trajectory slope as {\color{black}the} $\rho$ meson, we can construe two Regge trajectories for the $\omega$ meson {family}.  {Then} we also listed these observed $\omega$ states and $X(2240)$ for comparison with these values directly from the Regge trajectory analysis. We find
\begin{itemize}
\item $\omega(1420)$ and $\omega(1960)$ can be as the first and the second radial excitations of $\omega(782)$. There exist two $\omega$ states {\color{black}[$\omega(2290)$ and $\omega(2330)$]} as the third radial excitation of {$\omega(782)$}. $X(2240)$ as the third radial excitations of $\omega(782)$ is also possible.

\item  $\omega(2205)$ as the second radial excitation of $\omega(1650)$. We notice that the first radial excitation $\omega(2D)$  is still missing in experiment. Its mass is predicted to be 1940 MeV which is between the values {\color{black}1895  and 2179 MeV} given by Ref.  \cite{Ebert:2009ub} and Ref. \cite{Godfrey:1985xj}, respectively. { In fact, $\omega(1960)$ as the candidate of $\omega(2D)$ cannot be excluded by {analyzing the mass spectrum}. In the following, we  {will} also discuss this point.}\item  $\omega(2205)$ {can be assigned} as the second radial excitation of $\omega(1650)$. We notice that the first radial excitation $\omega(2D)$  is still missing in experiment. Its mass is predicted to be 1940 MeV which is between the values {\color{black}1895  and 2179 MeV} given by Ref.  \cite{Ebert:2009ub} and Ref. \cite{Godfrey:1985xj}, respectively.

\item{The analysis of Regge trajectory shows that the masses of $\omega(5S)$ and  $\omega(4D)$ are 2.42 GeV and 2.57 GeV, respectively. Here, the mass of $\omega(5S)$ is smaller than that given in Ref. \cite{Godfrey:1985xj} and   Ref. \cite{Ebert:2009ub},  which are 2.817 {GeV} and  2.472 GeV, respectively.}

\end{itemize}

\section{Getting the decay information}\label{s2}

For presenting the OZI allowed two-body decays of these discussed $\omega$ and $\omega$-like states, we employ
the QPC model in the concrete calculation{, where we assume that all decays are {to\//via} two-meson channels.} This model was {\color{black}first} proposed by Micu \cite{Micu:1968mk} and further developed by the Orsay group \cite{LeYaouanc:1972ae,LeYaouanc:1973xz,LeYaouanc:1974mr,LeYaouanc:1977gm,LeYaouanc:1977ux}. Until now,
the QPC model has been widely applied to study the OZI allowed two-body  strong decays of hadrons \cite{vanBeveren:1979bd,vanBeveren:1982qb,Capstick:1993kb,Page:1995rh,Titov:1995si,Ackleh:1996yt,Blundell:1996as,
Bonnaz:2001aj,Zhou:2004mw,Lu:2006ry,Zhang:2006yj,Luo:2009wu,Sun:2009tg,Liu:2009fe,Sun:2010pg,Rijken:2010zza,Ye:2012gu,
Wang:2012wa,He:2013ttg,Sun:2013qca,Pang:2014laa,Wang:2014sea,Chen:2015iqa,Pang:2017dlw,Pang:2018gcn}. Thus, the QPC model has become an effective {\color{black}way  to} illustrate the decay behavior of hadronic states. In the following, we will give a concise introduction to this model.

For a decay process $A\to B+C$, we may define the corresponding decay amplitude by
\begin{eqnarray}
\langle BC|\mathcal{T}|A \rangle = \delta ^3(\mathbf{P}_B+\mathbf{P}_C)\mathcal{M}^{{M}_{J_{A}}M_{J_{B}}M_{J_{C}}},
\end{eqnarray}
where {$\mathbf{P}_{B}$($\mathbf{P}_{C}$)} is the three-momentum of  {{the}}  meson $B(C)$ in the rest frame of  {{the}}  meson $A$. $M_{J_{i}}\, (i=A,B,C)$ denotes the
magnetic quantum number. The transition operator $\mathcal{T}$ describes a quark-antiquark pair creation from vacuum, which has quantum number
$J^{PC}=0^{++}$. The corresponding operator $\mathcal{T}$ satisfies
\begin{equation}
\begin{aligned}\mathcal{T}& = -3\gamma \sum_{m}\langle 1m;1~-m|00\rangle\int d \mathbf{p}_3d\mathbf{p}_4\delta^3 (\mathbf{p}_3+\mathbf{p}_4) \\
& ~\times \mathcal{Y}_{1m}\left(\frac{\textbf{p}_3-\mathbf{p}_4}{2}\right)\chi _{1,-m}^{34}\phi _{0}^{34}\left(\omega_{0}^{34}\right)_{ij}b_{3i}^{\dag}(\mathbf{p}_3)
d_{4j}^{\dag}(\mathbf{p}_4),
\end{aligned}
\label{eq:gamma}
\end{equation}
where the quark and antiquark are denoted by indices $3$ and $4$, respectively. $\gamma$ depicts the strength of the creation of $q\bar{q}$ from vacuum.
$\mathcal{Y}_{\ell m}(\mathbf{p})={|\mathbf{p}|^{\ell}}Y_{\ell
m}(\mathbf{p})$ are the solid harmonics. $\chi$, $\phi$, and $\omega$ denote the spin, flavor, and color wave functions, respectively.
The subindices $i$ and $j$ are the color of a $q\bar{q}$ pair.

The decay width reads as
\begin{eqnarray}
\Gamma&=&\frac{\pi}{4} \frac{|\mathbf{P}|}{m_A^2}\sum_{J,L}|\mathcal{M}^{JL}(\mathbf{P})|^2.
\end{eqnarray}
Here, $m_{A}$ is the mass of  {{the}}  initial state $A$.  {And}
the two decay amplitudes can related by the Jacob-Wick formula \cite{Jacob:1959at}, i.e.,
\begin{equation}
\begin{aligned}
\mathcal{M}^{JL}(\mathbf{P})&=&\frac{\sqrt{4\pi(2L+1)}}{2J_A+1}\sum_{M_{J_B}M_{J_C}}\langle L0;JM_{J_A}|J_AM_{J_A}\rangle \\
&&\times \langle J_BM_{J_B};J_CM_{J_C}|{J_A}M_{J_A}\rangle \mathcal{M}^{M_{J_{A}}M_{J_B}M_{J_C}}.
\end{aligned}\label{eq:ww
}\end{equation}

In Ref.  \cite{He:2013ttg}, {\color{black}the} Lanzhou group once systematically studied {\color{black}the} $\rho$ meson family. Here, $\gamma$ as the input parameter of {\color{black}the} QPC model is taken {\color{black}as} 8.7 for $u \bar u/d \bar d$ pair creation, while the strength
of the $s\bar s$ pair creation satisfies $\gamma=8.7/\sqrt{3}$ \cite{LeYaouanc:1977gm,Ye:2012gu,He:2013ttg}. In the present calculation, we still adopt the same $\gamma$ value as that suggested in Ref.  \cite{He:2013ttg}. An important consideration is due to the similarity between the discussed $\omega$ meson family and the $\rho$ meson family.

When performing the calculation of the spatial integral of the decay amplitude, we apply {\color{black}a} simple harmonic {\color{black}oscillator  wave} function $\psi_{n\ell m}({\bf k})=R_{n\ell}(R,{\bf k})\mathcal{Y}_{n\ell m}(\bf{k})$ to describe the meson wave function involved in the decays.
When reproducing the  root mean square radius by solving the Schr\"odinger equation with the effective potential \cite{Close:2005se},
the parameter $R$ of the final channel in the {\color{black}simple harmonic oscillator} wave function is determined. For these discussed $\omega$ and $\omega$-like states, we may set a $R$ range
between the  $R$ values determined by the potential model in Refs. \cite{Close:2005se,Godfrey:1985xj}.

With these {\color{black}preparations}, we give the decay information of  these discussed $\omega$ and $\omega$-like states. In the following discussion, we use
{ $\rho$ and $\omega$} {\color{black}instead} of $\rho(770)$ and $\omega(782)$ in the final decay channels, respectively.

\subsection{The $S$-wave $\omega$ states}

\subsubsection{$\omega(1420)$}

In Fig. \ref{omegapar}, we list the resonance parameter of $\omega(1420)$ from different experimental groups, which explicitly shows the inconsistence existing in the measurements of the  {mass and width of this {$\omega$ state}}.

\begin{figure}[!htbp]
\hspace{-00pt}
\begin{tabular}{ccc}
\scalebox{0.9}{\includegraphics{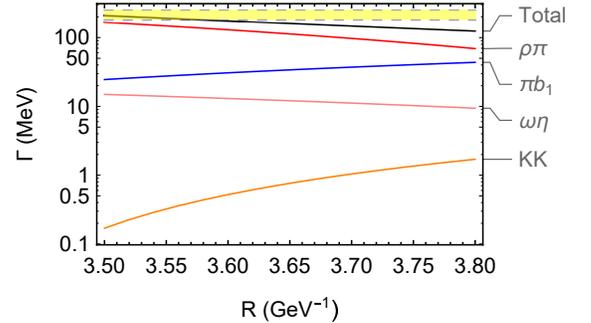}} \\
\end{tabular}
\caption {The $R$ dependence of the calculated partial and total decay widths of  $\omega(1420)$ as a $\omega(2S)$ {\color{black}state}. Here, the yellow band is the PDG estimate of the width of $\omega(1420)$ \cite{Tanabashi:2018oca}.}
 \label{fig:2S1425}
\end{figure}

As indicated in the Regge trajectory analysis shown in Fig. \ref{regge}, $\omega(1420)$ can be assigned as the second radial excitation of $\omega(782)$, which was discussed in {the} previous work \cite{Wang:2012wa,Barnes:1996ff}. In this work, we recalculate
the OZI allowed two-body decays of $\omega(1420)$, which {are} listed in Fig.~\ref{fig:2S1425}. Here, the partial and total decay widths dependent on $R$ value are given. When taking $R= (3.5  -  3.8)$ GeV$^{-1}$, the obtained total width of $\omega(1420)$ is in the range of {\color{black}$(120-200)$} MeV, which may support the PDG estimate for the width {\color{black}($180-250$ MeV)} \cite{Tanabashi:2018oca} and is a little bit larger than the result (378 MeV) in Ref. \cite{Barnes:1996ff}. Our result shows that the main decay channel of $\omega(1420)$ is $\rho\pi$, which has branching ratio $0.56 - 0.8$ comparable with the RVUE's experiment measurement ($\Gamma(\rho\pi)/\Gamma_{{\rm total}}=0.699\pm0.029$ \cite{Henner:2002iv}).  {Then}, $\omega(1420)\to \pi
b_1(1235)$ has {\color{black}an} obvious contribution to the total width as illustrated in our calculation. Since $b_1(1235)$ dominantly decays into $\omega\pi$ \cite{Tanabashi:2018oca}, we can understand why $\omega(1420)\to \pi b_1(1235)$ and $\omega(1420)\to \omega\pi\pi$ were {\color{black}first} found in experiment \cite{Henner:2002iv}. In addition, $\omega(1420)\to\omega\eta$ is sizable, which was reported in the experimental analysis of the $e^+e^-\to \omega\eta$ process \cite{CMD-3:2017tgb,Aubert:2007ym}. As the subordinate decay mode, $\omega(1420)\to K\bar{K}$ is still missing in experiment.

Although {\color{black}the} $\omega(1420)$ as {\color{black}the} $\omega(2S)$ state is  {\color{black}no in doubt, an } experimental study of $\omega(1420)$ with higher precision is {\color{black}needed,  especially} for its resonance parameter and branching ratios.

\subsubsection{$\omega(3S)$ and its candidate $\omega(1960)$}

\begin{figure}[htbp]
\hspace{-00pt}
\begin{tabular}{ccc}
\scalebox{0.9}{\includegraphics{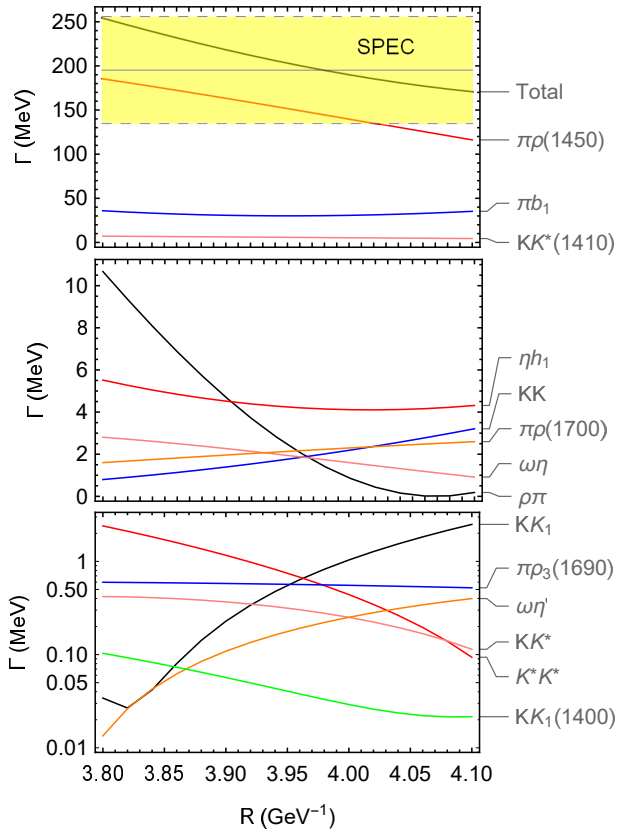}}\\
\end{tabular}
\caption {The calculated partial and total decay widths of $\omega(1960)$  dependent on $R$. Here, $\omega(1960)$ is treated as a $\omega(3S)$ state. We also list the experimental width of $\omega(1960)$  \cite{Anisovich:2011sva} for comparison with our result.  Here, we do not show these tiny modes which  have width below $0.1$ MeV. And, $b_1$ and $h_1$ represent the {\color{black}abbreviations} of $b_1(1235)$ and $h_1(1170)$, respectively.}
 \label{fig:3S1960}
\end{figure}

{\color{black}First}, we need to introduce the experimental status of $\omega(1960)$. Anisovich {\it et al.} performed a combined fit to the data of the $p\bar{p}\to \omega\eta, ~\omega\pi\pi$ processes, and found the $\omega(1960)$ signal with  {the mass of $m=1960\pm25$ MeV and the width of $\Gamma=195\pm60$ MeV}  \cite{Anisovich:2011sva}. Since there only exists one experimental result to $\omega(1960)$, in PDG $\omega(1960)$ is not collected into the particle list but as {a further state} \cite{Tanabashi:2018oca}.

According to the Regge trajectory analysis, it is suitable to recommend $\omega(1960)$ as a $\omega(3S)$ state. For further examining this  assignment, we study its decay behavior and find that the experimental width of $\omega(1960)$ can be reproduced when $R=(3.8-4.1)$ GeV$^{-1}$ as shown in Fig. \ref{fig:3S1960}. {Here} $\omega(1960)$ may mainly decay into $\pi \rho(1450)$, $\pi{b_1(1235)}$, and $\rho\pi$, which { have branch ratios $0.68-0.75$, $0.14-0.21$, and $0-0.04$, respectively, and} almost contribute the total width.  We want to specify that $\pi{b_1(1235)}$ can decay to $\omega\pi\pi$ which is the final channel adopted in the experiment analysis \cite{Anisovich:2011sva}.
In addition, its important decay {modes} are $\eta h_1(1170)$, $K\bar{K}$, $\omega\eta$, $\pi \rho(1700)$, in which $\omega\eta$ has been observed in experiment too  \cite{Anisovich:2011sva}. Besides, {the} other subordinate decay channels of $\omega(1960)$ can be found in Fig. \ref{fig:3S1960}.

In Fig. \ref{fig:3S1960}, we set a $R$ range to present that the result is dependent on $R$. This fact reflects that the node effect from the spatial wave function of {\color{black}the} initial state
is obvious especially when we discuss {the} higher radial excitations. It is the reason why we like to choose a $R$ range to  {discuss the} partial and total decay widths instead of taking {a  typical $R$ value}.

Further experimental confirmation  {of} $\omega(1960)$ is crucial to establish  {its observation}. Besides focusing on the precise measurement of its resonance parameter, we suggest {\color{black}experiments} to provide more abundant information of its partial widths.  {It may} stimulate {\color{black}a} theorist to promote the theoretical precision involved in the calculation of its decay behavior.

{We noticed that the mass of $\omega(1960)$ is close to the estimate mass of $\omega(2D)$.  {As shown} in Fig. {\color{black}\ref{fig:2D},that} the obtained total decay width of $\omega(2D)$ is a little bit larger than  {the central value of the mass} of $\omega(1960)$ ($\Gamma=195\pm60$ MeV) \cite{Anisovich:2011sva}. Thus, we cannot exclude the possibility of $\omega(1960)$ as the candidate of $\omega(2D)$ if {\color{black}we are} considering the experimental error. Here, further experimental  information of the ratio of $\pi b_1(1235)$ and $\pi\rho(1450)$ decay widths for $\omega(3S)$ and $\omega(2D)$ can be a key point when distinguishing these two possible assignments to $\omega(1960)$.}

\subsubsection{$\omega(4S)$ and three candidates $\omega(2290)$, $\omega(2330)$ and $X(2240)$}\label{x2240}
\begin{figure*}[htbp]
\hspace{0pt}
\begin{tabular}{cccc}
\scalebox{0.80}{\includegraphics{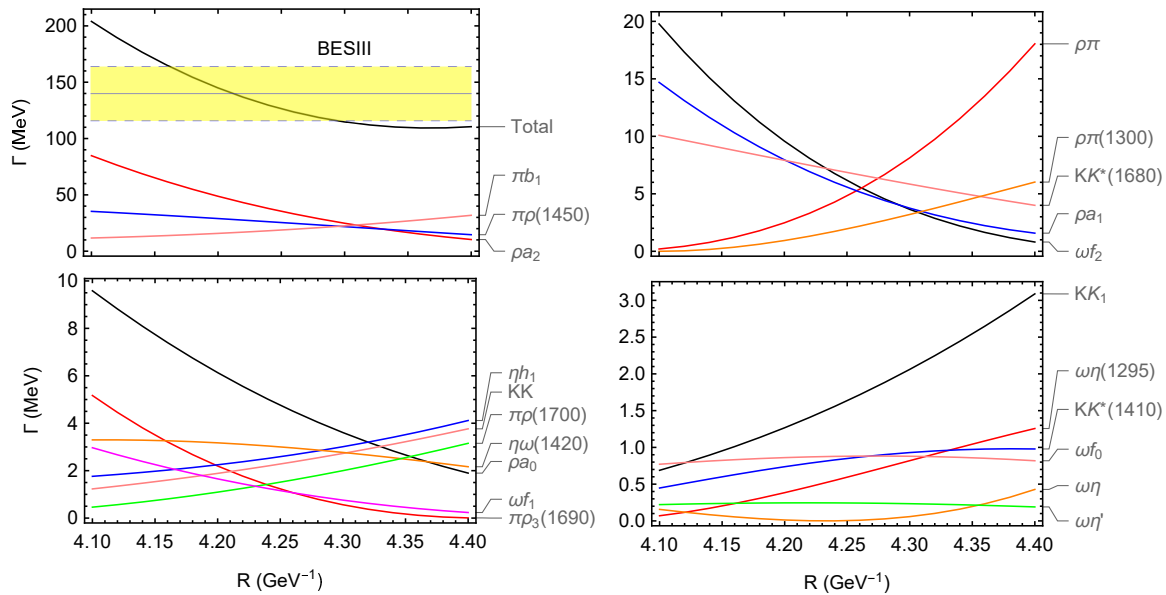}}\\
\end{tabular}
\caption {The $R$ dependence of the calculated partial and total decay widths of $\omega(4S)$. Here, we do not list these tiny decay mode contributions whose width are below 0.2 MeV. $b_1$, $h_1$, $a_0$, $a_1$, $a_2$, $f_0$, $f_1$,  $f_2$,  and $K_1$  represent $b_1(1235)$, $h_1(1170)$, $a_0(1450)$, $a_1(1260)$, $a_2(1320)$, $f_0(1370)$, $f_1(1285)$, $f_2(1270)$,  and $K_1(1270)$, respectively. {The band {\color{black}labeled}
BESIII is for the $X(2240)$ \cite{Ablikim:2018iyx}.}}
 \label{fig:4S}
\end{figure*}

As shown in Fig. \ref{regge}, there exist three possible candidates [$\omega(2290)$, $\omega(2330)$, and $X(2240)$]
for the $4^3\text{S}_1$ state in the $\omega$ meson family.
$\omega (2290)$ was found in the partial wave analysis of the data of $p\bar p\to \Lambda \bar{\Lambda}$ \cite{Bugg:2004rj} with the resonance {\color{black}parameters} $m=2290\pm20$ MeV and $\Gamma=275\pm35$ MeV. Another candidate of  $\omega(4S)$ is  $\omega(2330)$, which was observed in the process $\gamma p\to \rho^\pm\rho\pi^{\mp}$ by the  OMEG Collaboration \cite{Atkinson:1983yx}. The mass and width of $\omega(2330)$ are {$m=2330\pm30$ MeV and $\Gamma=435\pm75$ MeV,} respectively.  Similar to the case of $\omega(1960)$, these two $\omega$ states were collected into the further states of PDG \cite{Tanabashi:2018oca}  because {\color{black}there was} no further experimental confirmation.
In 2018, the BESIII Collaboration reported a $X(2240)$ in the $e^+e^- \to K^+K^-$ process, which has {the mass of $2239.2\pm 7.1\pm 11.3$ MeV and the width of} $139.8\pm12.3\pm20.6$ MeV \cite{Ablikim:2018iyx}. $X(2240)$ can be as the candidate of $\omega(4S)$ since {\color{black}it is possible that} $J^{PC}$, {\color{black}the  quantum number} is suitable and the analysis of {\color{black}the} Regge trajectory can support this scenario.
{Facing this} situation, it is necessary to study the decay behaviors of $\omega(4S)$, which may be applied  {\color{black}to} test the possibility of assigning these three candidates into {the} $\omega$ family. Here, we adopt the mass of $\omega(2290)$ as input to investigate the strong decay of $\omega(4S)$ since {\color{black}the} small mass difference of $\omega(4S)$ cannot change the conclusion.

According to the numerical result listed in Fig. \ref{fig:4S}, we may find that the width of $\omega (4S)$ is in the range of {\color{black}$(110-200)$ MeV}, which is consistent with the width of
$X(2240)$ and much smaller than that of $\omega(2290)$ and $\omega(2330)$. {Considering this result simply}, the assignment $\omega(4S)=X(2240)$ can be enforced. However, we still wait for further experimental information to make definite conclusion. Here, the experimental measurement of its partial decay widths is valuable .

Our result shows that $\omega(4S)$ mainly decays into $\rho a_2(1320)$, {\color{black}$\pi\rho(1450)$} and $\pi{b_1(1235)}$, which { have the branching ratios {of} $0.094-0.42$, $0.13-0.2$, and $0.057-0.29$, respectively}, and almost contribute the total width of $\omega(4S)$. {\color{black}Because of} $\rho \pi$ being {\color{black}the} dominant decay mode of $a_2(1320)$,  the three-body channel $\rho \rho \pi$ is an important decay mode of $\omega(4S)$. By the $\rho \rho \pi$ channel, $\omega(2330)$ was observed \cite{Atkinson:1983yx}.
Besides, $\rho a_0(1450)$, $\eta h_1(1170)$, and  $K\bar{K}$  { also} are the important decay modes. In experiment, $X(2240)$ was reported in the $K\bar{K}$ final state by BESIII \cite{Ablikim:2018iyx}. The detailed decay information can be found in Fig. \ref{fig:4S}.
It is obvious that establishing $\omega(4S)$ is still on the way.

\subsubsection{The predicted $\omega(5S)$}

In this work, we also study the $\omega(5S)$ state. $\omega(5S)$ has not been observed in experiment.
The Regge trajectory analysis shown in Fig. \ref{regge} indicates that $\omega(5S)$ has the mass of 2.57 GeV, which is {\color{black}like the} input when calculating the decay of $\omega(5S)$.

\begin{figure*}[!ht]
\hspace{-00pt}
\begin{tabular}{ccc}
\scalebox{0.8}{\includegraphics{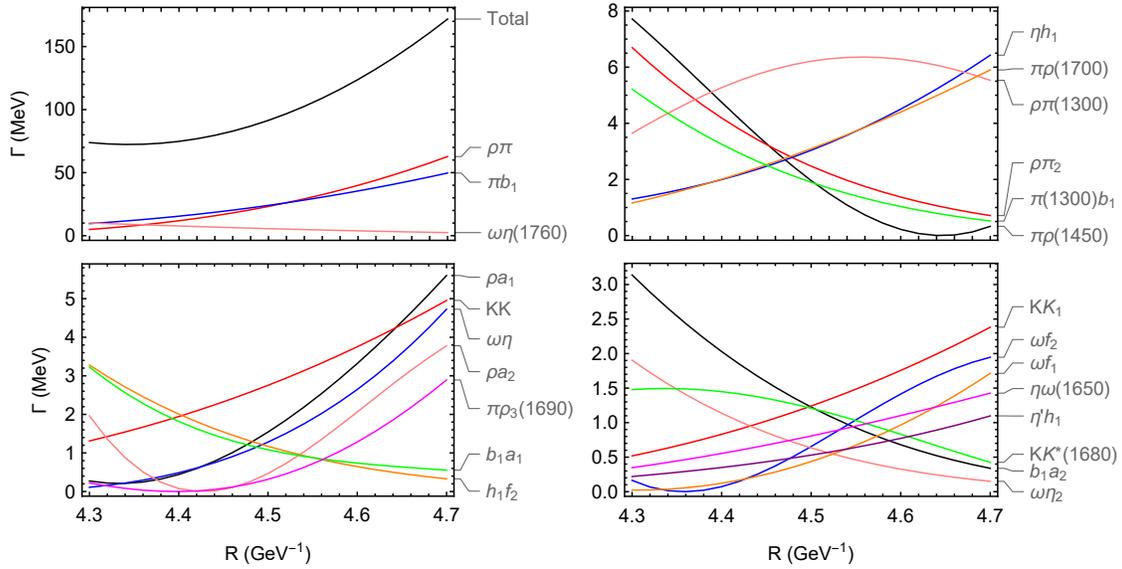}} \\
\end{tabular}
\caption {The $R$ dependence of the calculated partial and total decay widths of  $\omega(5S)$  state. Here, the tiny decay modes with  width bellow $1.0$ MeV are not listed. $b_1$, $h_1$, $a_0$, $a_1$, $a_2$, $f_0$, $f_1$, $f_2$,  $\eta_2$, $\pi_2$, and $K_1$  denote $b_1(1235)$, $h_1(1170)$, $a_0(1450)$, $a_1(1260)$, $a_2(1320)$, $f_0(1370)$, $f_1(1285)$, $f_2(1270)$, $\eta_2(1645)$, $\pi_2(1670)$, and $K_1(1270)$, respectively.}
 \label{fig:5S}
\end{figure*}

Here, the obtained OZI allowed two-body strong decay of $\omega(5S)$ by the QPC model is presented in Fig. \ref{fig:5S}.
The predicted $\omega(5S)$ has  {the width of} {\color{black}$(74-170)$} MeV which corresponds to the adopted parameter $R=(4.3-4.7) $ GeV$^{-1}$. Its
main decay modes are $\rho\pi$ and $\pi b_1(1235)$. Additionally,
the important decay modes are $\eta h_1(1170)$, $\pi\rho(1450)$, $\pi\rho(1700)$, $\rho \pi(1300)$, and $\rho \pi_2(1670)$.
 And, the $\rho a_1(1260)$, $K\bar{K}$, $\omega \eta$,  and $\rho a_2(1320)$ modes also have considerable contributions to the total decay width.
We hope that the predicted behavior of $\omega(5S)$ is helpful to {\color{black}the} experimental search for $\omega(5S)$. The running BESIII with {\color{black}a} higher precision has {\color{black}the} potential to capture the evidence of $\omega(5S)$.

\subsection{The $D$-wave $\omega$ states}

In this section, we focus on the analysis  {of} the $D$-wave $\omega$ mesons by combining with the experimental data of
$\omega(1650)$ and $\omega(2205)$. We also predict two missing {states:} $\omega(2D)$ and $\omega(4D)$.

\subsubsection{$\omega(1650)$}

 Although $\omega(1650)$ as a $D$-wave ground state of the $\omega$ meson family was suggested in {\color{black}the} literature  \cite{Wang:2012wa,Tanabashi:2018oca}, the measurement of its resonance parameter was not satisfactory as shown in Fig.  \ref{omegapar}.

 \begin{figure}[htbp]
\hspace{-00pt}
\begin{tabular}{ccc}
\scalebox{0.77}{\includegraphics{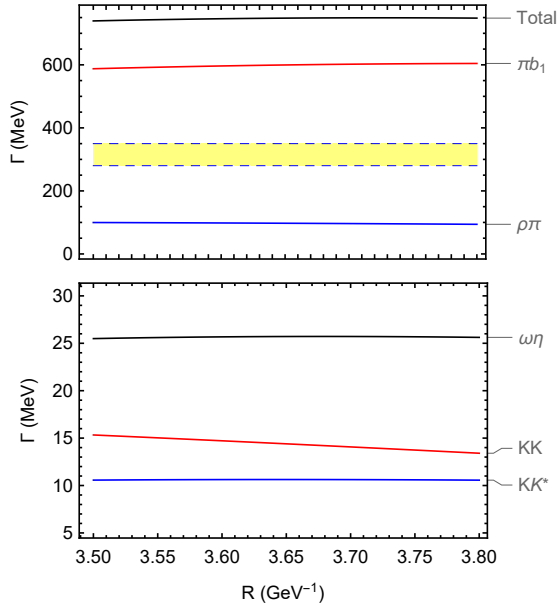}} \\
\end{tabular}
\caption {The $R$ dependence of the calculated partial and total decay widths of $\omega(1650)$ as a $1D$ state. Here, the yellow band {\color{black}is a PDG estimated} value of the width of $\omega(1650)$ \cite{Tanabashi:2018oca}.}
 \label{fig:1D}
\end{figure}

In this work, we calculate the partial and total decay widths dependent on {\color{black}the} $R$ value as shown in Fig. \ref{fig:1D}.
The obtained partial and total decay widths are not sensitive to the {\color{black}$R=(3.5-3.8)$} GeV$^{-1}$ range. {This can} be understood well since
there does not exist node effect. The total width of $\omega(1D)$ is about $750$ MeV, which is larger than the {\color{black}estimated} value  {of }$315\pm 35$ MeV in PDG \cite{Tanabashi:2018oca}.  {Further comparing} this theoretical result with other concrete experimental data, we find that
our result is close to the experimental width in Ref. \cite{Achasov:2016qvd}.

Among these calculated partial decay widths, the $\omega(1D)\to \pi b_1(1235)$ decay width as the dominant contribution to the total decay width is illustrated in our calculation, which is confirmed by the experimental data
${\Gamma_{\omega\pi\pi}}/{\Gamma_{Total}}=0.624\pm0.014$ \cite{Henner:2002iv}.
At the same time, we get  {that} the branching ratio of {\color{black}the} $\omega(1D)$ decay into $\rho\pi$ is $0.13$, {which is about {\color{black}3} times smaller than}
the {\color{black}experimental} value $0.38\pm 0.014$ \cite{Henner:2002iv}.
In addition, the branching ratios of the $\omega\eta$, {\color{black}$K\bar{K}$,  and} $K^*\bar{K^*}$ decay modes are sizable, which are about $0.033$, $0.02$, and $0.014$ respectively, in which $\omega\eta$  was observed in the $e^+e^-$ annihilation process \cite{CMD-3:2017tgb,Achasov:2016qvd}.

Generally, {it is suitable to explain $\omega(1650)$ as a $\omega(1D)$ state}. At present, a crucial task is to carry out the precise measurement of the resonance parameter of $\omega(1650)$ for {\color{black}clarifying the messy} situation of  the measured width of $\omega(1650)$.

\subsubsection{The predicted $\omega(2D)$}

As shown in Fig. \ref{regge}, we may construct a Regge trajectory for the $D$-wave $\omega$ mesons. We need to insert a $\omega(2D)$ state between $\omega(1650)$ and $\omega(2205)$. The $\omega(2D)$  state missing in experiment has {the mass of} 1940 MeV according to
the Regge trajectory analysis.

\begin{figure}[htbp]
\hspace{-00pt}
\begin{tabular}{ccc}
\scalebox{0.81}{\includegraphics{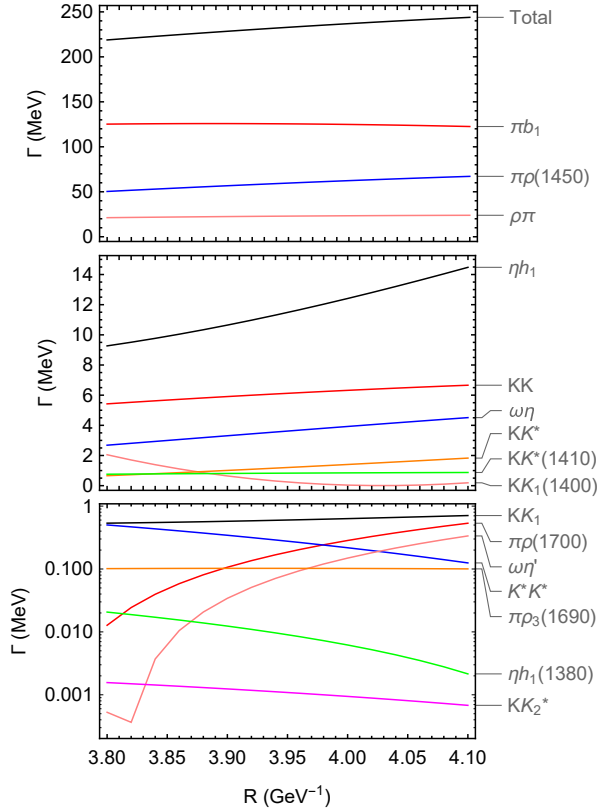}} \\
\end{tabular}
\caption {The $R$ dependence of the calculated partial and total decay widths of  $\omega(2D)$. Here, $b_1$, $h_1$, $K_1$ and $K_2^*$   represent $b_1(1235)$, $h_1(1170)$, $K_1(1270)$ and $K_2^*(1430)$, respectively.}
 \label{fig:2D}
\end{figure}

When further discussing the OZI allowed two-body decays of $\omega(2D)$ (see Fig. \ref{fig:2D} for more details),  we may estimate the total decay width of $\omega(2D)$ to be {\color{black}$(220-245)$} MeV. And the predicted main decay channels of $\omega(2D)$ include $\pi b_1(1235)$, $\pi \rho(1450)$, and $\rho\pi$, which are not sensitive to the $R$ value.
Since our calculation shows {\color{black}that} the branch ratio of $\pi b_1(1235)$ {\color{black}can  reach} up to $0.5$, a three-body decay $\omega(2D)\to \omega\pi\pi$  is suggested {to be an} ideal channel of searching for $\omega(2D)$. Additionally, the decay channels $\pi b_1(1235)$ and  $\rho\pi$ have branching ratios around $0.25$ and $0.1$, respectively. {Also, $\eta h_1(1170)$, $K\bar{K}$, and $\omega\eta$ are  important decay channels for this predicted $\omega(2D)$ state}.

\subsubsection{$\omega(3D)$ and its candidate $\omega(2205)$}

\begin{figure*}[!htbp]
\hspace{-20pt}
\begin{tabular}{ccc}
\scalebox{0.8}{\includegraphics{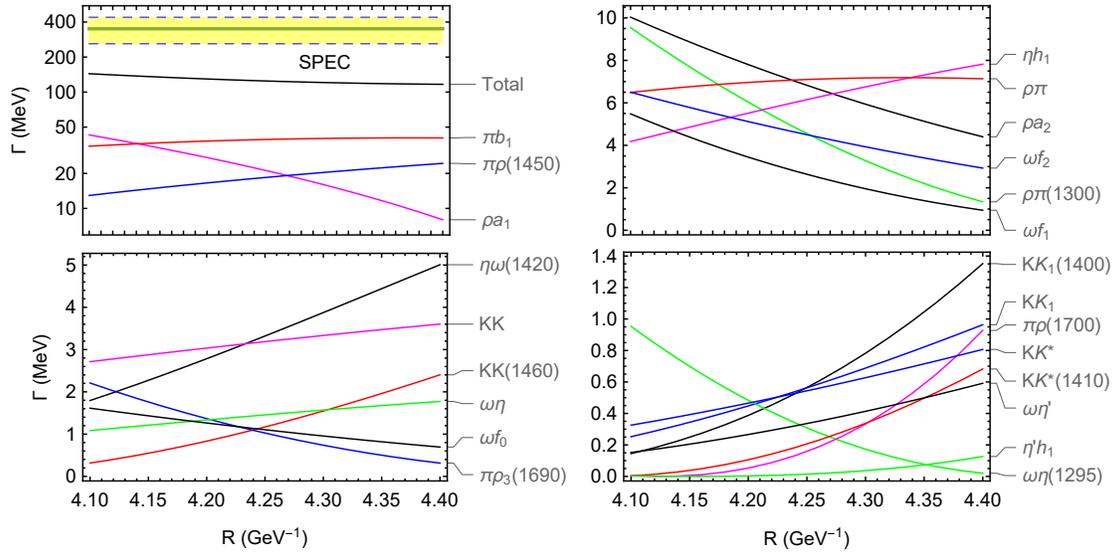}}\\
\end{tabular}
\caption {The $R$ dependence of the calculated partial and total decay widths of $\omega(2205)$ as $\omega(3D)$. Here, the yellow band is the experiment value from Ref. \cite{Anisovich:2011sva} and  these tiny decay modes with {\color{black}a} partial width lower than 1 MeV are not listed. $b_1$, $h_1$,  $a_1$, $a_2$, $f_0$, $f_1$, $f_2$,  and $K_1$ are the abbreviations of $b_1(1235)$, $h_1(1170)$,  $a_1(1260)$, $a_2(1320)$, $f_0(1370)$, $f_1(1285)$, $f_2(1270)$,  and $K_1(1270)$, respectively.}
 \label{fig:3D}
\end{figure*}

In the $p\bar{p}\to \omega\eta, ~\omega\pi\pi$ process, SPEC observed the $\omega(2205)$ state associated with $\omega(1960)$ \cite{Anisovich:2011sva}.
By analyzing the Regge trajectory of $\omega$ states, we can naturally assign $\omega(2205)$ as a $\omega(3D)$ state. The calculated  two-body strong decay information is collected into  Fig. \ref{fig:3D}.
By analyzing the Regge trajectory of {the} $\omega$ states, we can naturally assign $\omega(2205)$ as a $\omega(3D)$ state. The calculated  two-body strong decay information is collected into  Fig. \ref{fig:3D}.

The calculated total decay width of  $\omega(3D)$  is {\color{black}$(120-140)$} MeV when taking $R =(4.1-4.4)$ GeV$^{-1}$, which is smaller than the  experimental width value $\Gamma=350\pm90$ MeV \cite{Anisovich:2011sva}. We find that $\pi{b_1(1235)}$, $\rho{a_1}(1260)$, and $\pi\rho(1450)$  are the main decay modes of $\omega(3D)$, {which have the branch ratios {of} $0.24-0.35$, $0.069-0.3$, and $0.09-0.21$, respectively}. $\omega\pi$ is  {the} dominant decay mode of $b_1(1235)$ \cite{Tanabashi:2018oca}, which can explain that $\omega(2205)$ was observed in the
      $p\bar{p}\to  \omega\pi\pi$ process at the SPEC experiment \cite{Anisovich:2011sva}.
       { Since $\rho(1450)$ can decay into $\pi\pi$, $\pi\pi\pi$ will be {an  important} three-body final state of exploring $\omega(3D)$}.
Our result shows that $\rho a_2(1320)$, $\eta h_1(1170)$, $\rho \pi$, $\omega f_2(1270)$, {\color{black}$\rho\pi(1300)$}, and $\omega f_1(1285)$ are  its  subordinate decay channels in which the branch ratio of $\rho a_2(1320)$ is {\color{black}$0.04-0.07$}.
Among these sizable decay channels  {of}
      $K\bar{K}$, $\omega(1420)\eta$,  {\color{black}{$K(1460)\bar{K}$+H.c.}}, and $\omega\eta$, only $\omega\eta$ is  observed in experiment  \cite{Anisovich:2011sva}. The details of other decay channels can be found by Fig. \ref{fig:3D}.

{In Sec. \ref{x2240}, we  {discussed} the possibility of $X(2240)$ as $\omega(4S)$. In fact, $X(2240)$ as a $\omega(3D)$ state can be put into the Regge trajectory of the {\color{black}D-wave} $\omega$ mesons. The obtained total decay width of $\omega(3D)$ overlaps with
 the experimental width of $X(2240)$.  {{It will be a crucial task to distinguish the $\omega(4S)$ and $\omega(3D)$ assignments to $X(2240)$ },}
 since the decay behavior of $\omega(4S)$ is similar to that of $\omega(3D)$ as shown in {\color{black}Figs. \ref{fig:4S} and \ref{fig:3D}}. We may notice that there exists {\color{black}an} obvious difference of the branching {ratios of $\rho a_2(1320)$ and $\rho a_1(1260)$} decay channels for $\omega(4S)$ and $\omega(3D)$. Thus, the {\color{black}measurements} of  {$\rho a_2(1320)$ and $\rho a_1(1260)$} {\color{black} can} be applied to distinguish the above assignment to $X(2240)$.}

Obviously, the present experimental data is still not enough  {to establish} the $\omega(3D)$ state.

\subsubsection{The predicted $\omega(4D)$}

\begin{figure*}[htbp]
\hspace{-20pt}
\scalebox{0.8}{\includegraphics{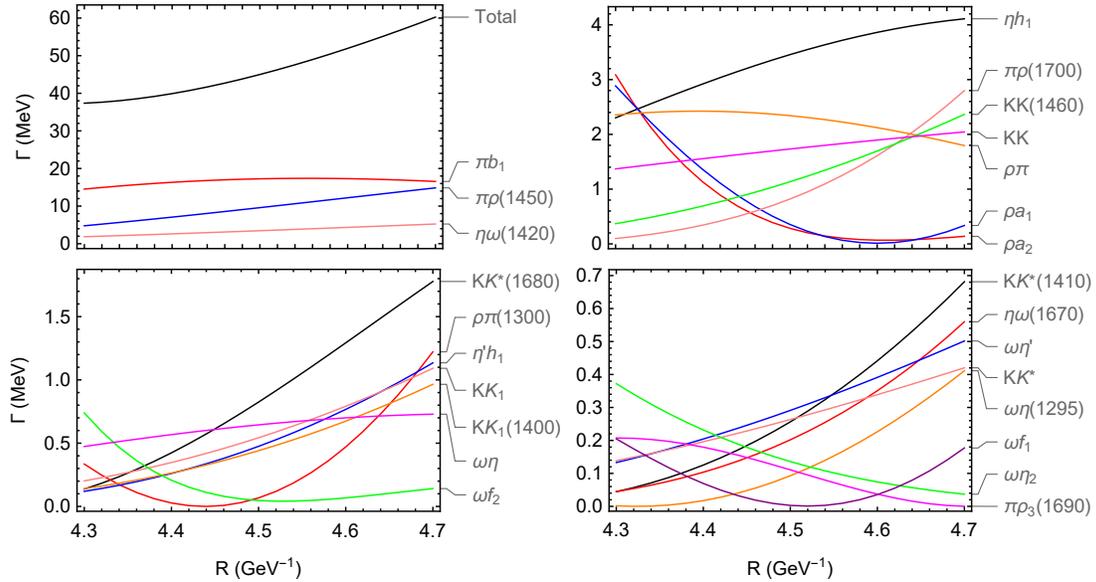}}\\
\caption {The $R$ dependence of the calculated partial and total decay widths of $\omega(4D)$. Here, these partial decay widths with width smaller than 0.2 MeV are not listed. And, $b_1$, $h_1$,   $a_1$, $a_2$, $f_0$, $f_1$, $f_2$,  $\eta_2$,  and $K_1$   represent $b_1(1235)$, $h_1(1170)$,  $a_1(1260)$, $a_2(1320)$, $f_1(1285)$, $f_2(1270)$, $\eta_2(1645)$,  and $K_1(1270)$, respectively.}
 \label{fig:4D}
\end{figure*}

 {\color{black}In the following}, we want to present our result of the predicted $4^3\text{D}_1$ state in the $\omega$ meson family.
By the analysis of Regge trajectory (see Fig. \ref{regge}), we predict that
$\omega{(4D)}$ {\color{black}has a mass of} about $2420$ MeV. With such mass estimate for $\omega(4D)$, we calculate the partial and total decay widths of  $\omega(4D)$ which are collected into Fig. \ref{fig:4D} as the $R$ dependence.

One notices that $\omega(4D)$ is a narrow state with the total decay width {\color{black}of around (40-60) }MeV, where we take $R=(4.3-4.7)$ GeV$^{-1}$.
$\omega(4D)$ mainly decays into  $\pi{b_1(1235)}$, $\rho(1450)\pi$,  and $\omega(1420)\pi$. The branching ratios of these three typical decay modes are $0.27-0.49$, $0.13-0.25$, and $0.05-0.09$, respectively. $\eta{h_1(1170)}$, $\rho\pi$, $K\bar K$, {\color{black}$K\bar{K}(1460)$+H.c}.,  $\rho a_1(1260)$, and  $\rho a_2(1320)$ are the important decay modes of $\omega(4D)$. Here, we suggest $K\bar K$ as an ideal channel to search for the predicted $\omega(4D)$ state in experiment according to the last BESIII experience of studying light vector states in Ref. \cite{Ablikim:2018iyx}.

In this work,  {\color{black}{$K\bar{K^*}(1680)$+H.c.}}, {\color{black}$\rho\pi(1300)$,  $\eta^\prime h_1(1170)$  }etc.  are the subordinate decay channels.
At present, we can only provide this decay information for further experimental exploration of this higher $\omega$ state.

\section{SUMMARY}\label{s3}

Studying light hadron spectroscopy is an interesting research topic.
When checking the experimental data of these reported $\omega$ states, we notice two serious problems{\color{black}.} (1) The difference of the measured resonance parameters for the same state from different experimental groups is obvious.
Here, $\omega(1420)$ and $\omega(1670)$ as lower states are typical examples; (2) For the remaining higher states like $\omega(1960)$, $\omega(2290)$, $\omega(2330)$, and $\omega(2205)$, the corresponding experimental information is scarce.
It {\color{black}means} that these $\omega$ states are collected by PDG as further states.
Considering the {\color{black}messy situation of measurements} of these observed $\omega$ states listed in PDG \cite{Tanabashi:2018oca} and inspired by the last {\color{black}BESIII} observation of $\omega$-like state $X(2240)$ in the $e^+e^-\to K^+
{K}^-$ process \cite{Ablikim:2018iyx}, we perform a systematic study of the mass spectrum and the two-body OZI allowed strong decay of  the $\omega$ meson family.

 By comparing the experimental data with our theoretical result, we examine {\color{black}the} possibilities of assigning these reported $\omega$ states and {\color{black}the} $\omega$-like state into the $\omega$ meson family. In addition, we predict the masses and decay behaviors of $\omega(5S)$, $\omega(2D)$ and $\omega(4D)$, which are still absent in experiment.
 Obviously, exploring them will be a key step to construct {\color{black}the} whole $\omega$ meson family. Our theoretical results can help {\color{black}an} experimentalist to select {\color{black}a} suitable channel {and find }these missing states.

{Recently}, the BESIII Collaboration announced {\color{black}the} { ``{\it White Paper on the Future Physics Programme of BESIII} '' \cite{Ablikim:2019hff}}, where investigating the light
hadron spectroscopy will still be one of the most important goals of the BESIII experiment. {\color{black}In  past years}, BESIII has paid more efforts to it.
Among {\color{black}the} { direct production of light vector states via the $e^+e^-$ annihilation results in the observation of some enhancement  structures around 2 GeV \cite{Ablikim:2010au,Ablikim:2014pfc,Ablikim:2018xuz,Ablikim:2018iyx},
$X(2240)$ discussed in this work is a typical state}. Thus, for BESIII, exploring {the}  $\omega$ meson states produced directly through  the $e^+e^-$ annihilation will be a research issue full of opportunity. We believe that the present work is a new {\color{black}starting} point of studying {the} $\omega$ meson spectroscopy.

\section{ACKNOWLEDGMENTS}
This work is supported  by the National Natural Science Foundation of China under Grants No. 11965016 and No. 11861051, the projects funded by Science and Technology Department of Qinghai Province (No. 2018-ZJ-971Q and No. 2019-ZJ-A10), the Key Laboratory of IoT of Qinghai under {\color{black}(Grant No. 2020-ZJ-Y16)}, and the Natural Science Foundation of Ningxia{\color{black}(Grant No. 2018AAC03063)}. {\color{black}X.L.} is supported by the China National Funds for Distinguished Young Scientists under Grant No. 11825503, the National Program for Support of Top-notch Young Professionals, and the projects funded by Science and Technology Department of Qinghai Province No. 2020-ZJ-728.

\bibliographystyle{apsrev4-1}
\bibliography{hepref}
\end{document}